\documentclass[12pt]{article}
\usepackage{epsfig} \usepackage{amsmath} \usepackage{amssymb}
 \usepackage{pifont} \usepackage{psfrag} 
  \def\ie{\hbox{\it i.e.}{ }} \def\etc{\hbox{\it etc.}{ }}   
  \def\eg{\hbox{\it e.g.}{ }}      
\newcommand{\ns}{\normalsize}
\newcommand{\beq}{\begin{equation}}
\newcommand{\eeq}{\end{equation}}                              
\def\eV{\relax\ifmmode{\rm e\kern-0.12em V}\else{\rm e\kern-0.12em V}\fi}
\def\MeV{\relax\ifmmode{\rm M\eV}\else{\rm M\eV}\fi}
\def\GeV{\relax\ifmmode{\rm G}\eV\else{\rm G\eV{ }}\fi}        
\def\al{\relax\ifmmode\alpha\else{$\alpha${ }}\fi}
\def\as{\relax\ifmmode \alpha_s\else{$ \alpha_s${ }}\fi}       
\def\alps{\relax\ifmmode\alpha_s\else{$\alpha_s${ }}\fi}
\def\albars{\relax\ifmmode{\bar{\alpha}_s}\else{$\bar{\alpha}_s${ }}\fi}
 \def\Agoth{\relax\ifmmode{\mathfrak A}\else{${\mathfrak A}${ }}\fi}
 \def\Agothk{\relax\ifmmode{\mathfrak A}_k\else{${\mathfrak A}_k${ }}\fi}
 \def\Acal{\relax\ifmmode{\cal A} \else{${\cal A} $ { }}\fi}
  \def\Acalk{\relax\ifmmode{{\cal A}_k}\else{${\cal A}_k$ { }}\fi}
\begin{document}
\begin{center}
 {\large\bf Novel Sets of Coupling Expansion \\
               Parameters for low-energy pQCD\footnote{A preliminary
 version with the main results was published in Ref.\cite{parus07}.}\\
 \bigskip
 {\ns D.~V.~Shirkov} \\
 \medskip

 {\it\normalsize Joint Institute for Nuclear Research,
  Dubna, 141980, Russia}}
\end{center}

 \begin{abstract}
 In quantum theory, physical amplitudes are usually presented in the
 form of Feynman perturbation series in powers of coupling constant
 $\al\,.$ However, it is known that these amplitudes are
 not regular functions at $\alpha=0\,.$ \par
  For QCD, we propose new sets of expansion parameters
 ${\bf w}_k(\as)$ that reflect singularity at $\as=0\,$ and should
 be used instead of powers $\as^k.$ Their explicit form is motivated
 by the so called Analytic Perturbation Theory. These parameters
 reveal saturation in a strong coupling case at the level \
 $\as^{eff}(\as\gg1)={\bf w}_1(\as\gg 1) \sim 0.5\,.$ They can be used
 for quanitative analysis of divers low-energy amplitudes. \par
   We argue that this new picture with non-power sets of perturbation
 expansion parameters, as well as the saturation feature, is of a
 rather general nature.
 \end{abstract}

 \section{\large Subject and Motivation}
 In quantum theory, physical amplitudes are usually presented in the
 form of perturbative series in powers of an expansion parameter
 $g\,$ related to intensity of interaction, the non-linear term
 in the equation of motion. As it is known from the early 50s
 \cite{dyson52}, these amplitudes are not regular functions at
 $\alpha=0\,,$ irrelevant to the existence of UV divergencies and
 renormalization. The most general and transparent argument
 \cite{Dingle73} can be formulated via a representation of path
 integral.

  In QFT, one deals with Feynman perturbation theory (PT) series in
 powers of a numerical parameter \al. In particular, in current
 practice, such a series
 for a QCD observable serves as a launch pad for Renorm-Group (RG)
 invariant expansion in powers of invariant/effective coupling
 $\albars(Q^2)\,$ or $\albars(s).$ As a rule, this function \albars,
 being a sum of ultraviolet (UV) logs, obeys phantom singularities,
 like  the so called Landau pole. In QCD, being located at a scale
 of a few hundred \MeV, they spoil the low-energy applications.\par

  From the mathematical point of view, a possibility of convergent
 power expansion implies that the expanded function $f(\al)\,$ is a
 regular (analytic) function of its argument at small \al. Meanwhile,
 it is known for sure\cite{dyson52} (also
 Refs.\cite{Sh:76lmp,Sh:77lnc}) that in the complex \al plane,
 there is an essential singularity at the origin $\al=0.$
 Correspondingly, at a small real positive \al, the PT
 series $\sum_n^{\infty}c_n\al^n\,$ is divergent with $c_n\sim n!\,$
 \ at \ $n\gg 1\,.$ Nevertheless, under some condition, a finite sum
 $\sum_{n}^{N}c_n\al^n\,$ can serve as a means for numerical
 approximation of the expanded function $f(\al)\,.$ \par

 In such a situation, the value \ $n^*\sim 1/\al\,$ is a critical one.
 Here, PT expansion can start to explode. Indeed, in the low-energy
 QCD, at \ $\as\sim 0.2-0.3\,,\,\, n^*\sim 3-5\,$ and thus the value
 of the 3- and 4-loop calculation is under question. Examples are
 known (see,\eg Table 2 in refs.\cite{Sh01epjc} and \cite{ShSol06}).\par

  How serious is this menace for practical low-energy pQCD
 calculation? Is it possible to use some other expansion parameter
 ${\bf w}(\al)\,$ instead of \al or a non-power set
 $\{{\bf w}_k(\al)\}\,$ of parameters\footnote{As it was proposed,
 \eg, by Caprini and Fischer \cite{cf00}.} instead of $\{\al^k\}$?

   Below, we try to answer this question by using a combination of
 rather general arguments including the principles of causality and
 renormalizability as well as self-consistency condition of theoretical
 description with respect to conversion from one physical picture
 (representation) to another by a suitable integral transformation.\par

  During the last decade, on the basis of these principles, a special
 scheme for ghost-free calculations in QCD was proposed \cite{apt96-7}
 and elaborated\cite{apt98-9}. It is known now as Analytic
 Perturbation Theory (APT). For fresh reviews, we recommend
 Refs.\cite{ShSol06,apt-revs}. In what follows, we shall use the APT
 notation and results. Compendium of a few relevant APT definitions
 is presented below in Appendix A. \par
    Discussing, in Conclusion, the possible meaning of our particular
 results, we involve additional evidence from the soluble
 QFT models with an infrared fixed point.

  \section{\bf\large Equivalence of Transformations}  

 Within the Analytic Perturbation Theory, a transition from the common
 QCD effective coupling function $\as(L)\,,\,L=\ln(Q^2/\Lambda^2)\,$
 to the Euclidean $\al_E(L)\,$ or Minkowskian $\al_M(L)\,$ one \etc
 can be treated as a transition to a new expansion parameter. For
 example, in the 1-loop case, at $\as\,, L > 0\,$ the conversion
 \[\as\to\al_M(\as) = {\bf w}^M(\al)=
 \frac{\arctan(\pi\beta_0\,\as)}{\pi\,\beta_0}\sim
 \as - \frac{\pi^2\beta_0^2}{3}\,\as^3+\dots\,\]
 induces a transition to the new effective coupling
 \[\alps(L)=\frac{1}{\beta_0\, L}\to\al_M(L)=
 \frac{\arctan\left(\pi/L\right)}{\pi\,\beta_0}\sim
 \alps(L)-\frac{\pi^2\beta_0^2}{3}\left[\alps(L)\right]^3
 +\dots\,\,.\]

  Generally, all the APT non-power expansion functions,
 Minkowskian ${\Agothk(s),\,}$ Euclidean ${\Acalk(Q^2)\,}$
 or Distance ${\aleph_k(r^{-2})\,}$, are mapped via the
 relation
\begin{equation}\label{map}
 L\to\Phi(\as)=-\int^{\as}\frac{d\,a}{\beta(a)}\,=\frac{1}
 {\beta_0\as}+\frac{\beta_1}{\beta_0^2}\,\ln\left(\frac{
 \beta_1}{\beta_0}+\frac{1}{\as}\right)+ O(\beta_2)\eeq
 $L=\ln(s/\Lambda^2)\,,\ln(Q^2/\Lambda^2)\,,
 \ln(1/r^2\Lambda^2)\,,$ on sets \
 $\{{\bf w}^{\rm APT=M,E,D}_k(\as)\}\,$
 {\[\Agothk\,\to\,{\bf w}^M_k(\as),\,\quad\Acalk\,\to
 {\bf w}^E_k(\as),\,\quad\aleph_k\,\to{\bf w}^D_k(\as)
 \,.\]}\vspace{-4mm}

  The functions ${\bf w}_k^{APT}\,$ resultant from different
 sources are related by integral transformations that stem
 from ones connecting the ``parent" APT expansion functions.
 For example, from Adler and Fourier transformations
 \[\Acalk(Q^2)=
 Q^2\int^\infty_0\frac{d\,s\,\Agothk(s)}{(s+Q^2)^2}\,,\quad
  \aleph_k(r)= r\,\int^\infty_0\,d Q\,\sin(Qr)\,\Acalk(Q^2)\]
 and eq.(\ref{map}) there follows a general relation
\begin{equation}\label{trans}
 {\bf w}_k^E(\al)= \int^\infty_0 d\,a\,K_{E\,Z}(\al,a)\,
                      {\bf w}_k^Z(a);\quad Z=M, D\,.\eeq

 \subsection{The one-loop case}     
 In the one-loop case, consequent elements are connected
 by simple recurrent relation
 \beq\label{3}
 {\bf w}^{(1)}_{k+1}(\al)= \frac{\al^2}{k}\,
\frac{d\,{\bf w}^{(1)}_k(\al)}{d\,\al}\,,\eeq
 while eq.(\ref{trans}) for $Z=M\,$ takes the form
\begin{equation}\label{trans1}                       
 {\bf w}_k^E(\al)=\frac{1}{2}\,\int^\infty_{-\infty}
 \frac{d\,a}{a^2}\,\frac{{\bf w}_k^M(a)}
 {1+\cosh\left(1/\beta_0\al-1/\beta_0\,a\right)}\,. \eeq

 The novel ``Minkowskian" and ``Euclidean"
 one-loop expansion functions
 \begin{equation}\label{w1}                         
 \al^{(1)}_{M}(\al)={\bf w}_1^{M,(1)}=
 \frac{\arctan(\pi\beta_0 \,
 \al)} {\pi\,\beta_0} \,;\quad {\bf w}_1^{E,(1)}(\al)=\al
 +\frac{\beta_0^{-1}} {1-e^{1/(\beta_0\al)}}\,;\eeq

 \beq \label{w2}{\bf w}_2^{M,(1)}=                  
 \frac{\al^2}{1+\pi^2\beta_0^2\,\al^2}\,;\quad
 {\bf w}_2^{E,(1)}(\al)= \al^2 -
 \frac{\beta_0^{-2}\,e^{1/(\beta_0\al)}} 
  {[1-e^{1/(\beta_0\al)}]^2}\,;\,   \eeq

  \beq \label{w34}                                  
 {\bf w}_3^{M,(1)}(\al)=\frac{\al^3}{[1+(\pi\beta_0
 \,\al)^2]^2} \,;\quad{\bf w}_4^{M,(1)}(\al)=\frac{\al^4\,
 (3-\pi^2\,\beta_0^2\,\al^2)} {3\,[1+(\pi\beta_0\,\al)^2]^3}\,
 ;\eeq
 are mutually connected by relations (\ref{3}),
 (\ref{trans1}). \smallskip

   All the functions  ${\bf w}_k^{\cdots}(\as)\ \mbox{as}
 \ \as\to\infty$ have finite limits
 {\small
 \[{\bf w}_1^M(\infty)=\al_M(\infty)={\bf w}_1^E(\infty)=
 \al_E(\infty)=\frac{1}{2\,\beta_0} \sim 0.7\,;\quad
 {\bf w}_2^{M,(1)}(\infty)=\frac{1}{\pi^2\beta_0^2}
 \,\sim 0.20\,;\]
 \[ \quad{\bf w}_2^{E,(1)}(\infty)=
 \frac{1}{12\beta_0^2}\,\sim 0.16\,;\quad 
{\bf w}_3^{M,(1)}(\infty)=0\,;\qquad{\bf w}_4^{M,(1)}(\al)
 =- \frac{1}{(\pi\,\beta_0)^4}\sim - 0.04\,. \] }

  Here, the limit $\bf{\as\to\infty},$ by (1), is
 adequate to ${L\to +0}.$ Curious enough, a physical region
 below $\Lambda\,$ (\ie, ${L< 0}\,$) corresponds to
 negative \as values. \vspace{-2mm}

  \subsection{Some properties of functions
 ${\bf w}^{\rm APT}_k(\as)\,$}                
 A few \underline{general properties} of novel expansion
 functions are of interest
\begin{itemize}\itemsep -1mm
\item Like their APT ``parents" \
 $\Agothk,\,\Acalk,\,\aleph_k\,,$  functions
 ${\{\bf w}^{\cdots}_k(\as)\}$ \ {\it in the whole real
 positive domain $(-\infty\,<\as <+\infty)$} \ form
 non-power sets of oscillating functions with $k\,$ zeroes.
\item Natural scales for them are  {$\al_M^*=
 \tfrac{1}{\pi\beta_0}\sim {\ns 0.5}\,,\, \al_E^*=
 \tfrac{1}{\beta_0}\sim {\ns 1.4}\,.$ }
\item Some of the functions, like ${\bf w}_k^E\,,$ obey
 singularity \ ${e^{1/(\beta_0\, \as)}}\,.$
\item They are not sensitive to ``their family origin". In
 Fig.1, curves ${\bf w}^E_{1,2}$ \ are close to \
 ${\bf w}^M_{1,2}\,$ (as compared \ to the Caprini-Fischer
 curves  ${w^{cf}_{1,2}}\,$ obtained\cite{cf00} by conformal
 transformation of the  Borel image). As can be shown, the
 same is true for ${\bf w}^D_{1,2}\,.$
\end{itemize}\vspace{-5mm}  

\begin{figure}[th]
 \unitlength=1mm
 \begin{picture}(150,100)
  \put(-20,-120){\epsfig{file=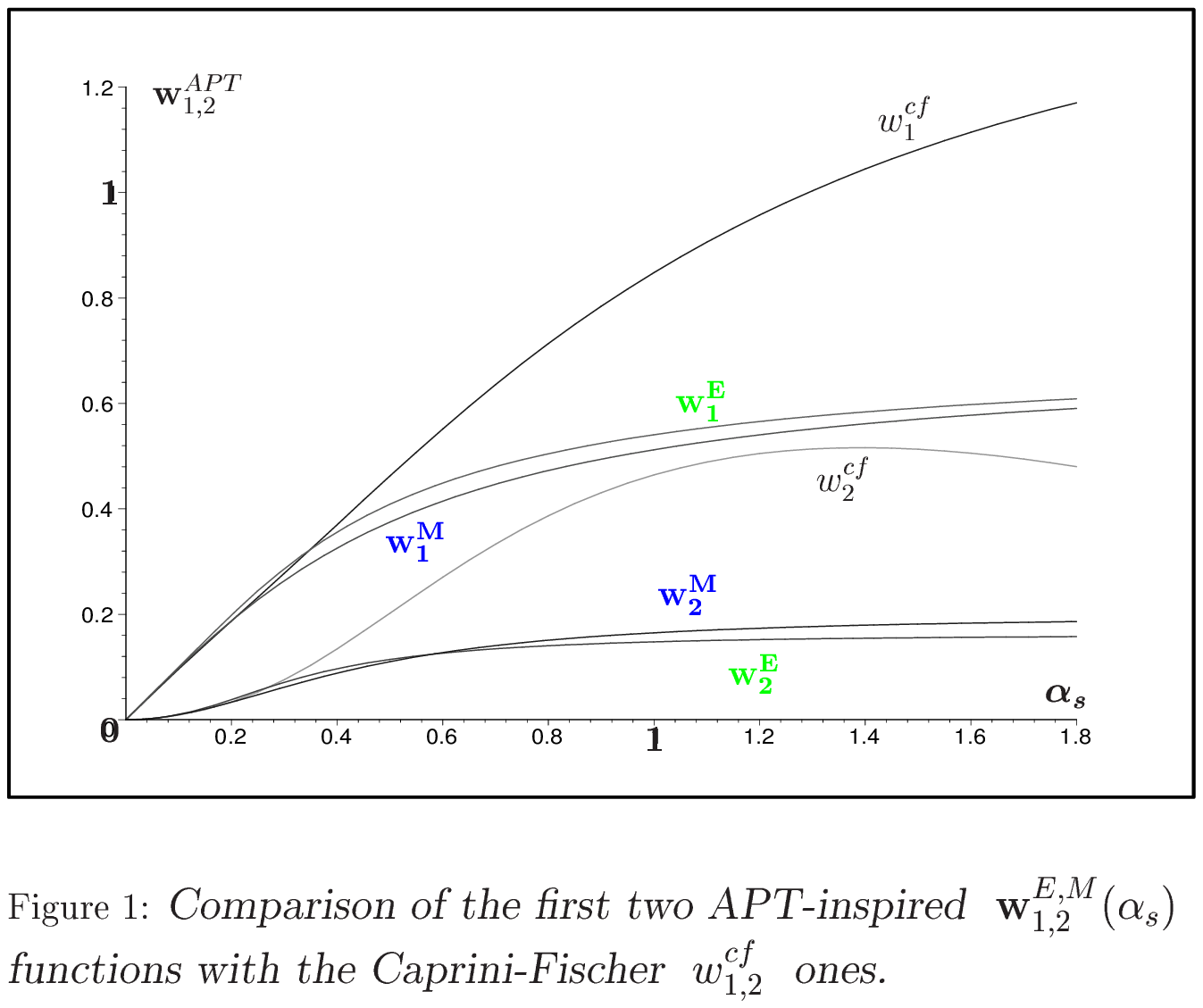,width=18cm}}
\end{picture}
\end{figure} 

  Some \underline{qualitative properties} of
  ${\bf w}^{\rm\small APT}_k(\as)$ should be mentioned
    \parindent 6mm

 1. APT-inspired functions ${\bf w}^{\rm APT}_{1,2}(\as)\,$
 deviate from powers $\as, \as^2\,$ at $\as\sim 0.3-0.4\,,$
 which corresponds to a few \GeV region.\par
 2. Quick {saturation} of the first {${\bf w}_1(\as)\,$} at
 0.4 and second ${\bf w}_2(\as)\,$ at $\sim$ \ 0.15 values.
  \parindent 6mm

  Thus, ``strong coupling" means ${\al_{eff}\lesssim 0.5}\,$.
 Physically, due to this, in a few GeV region effective
 QCD coupling should be less than 0.5.\par

  3. Relative difference between functions ``of various origin"
 is small (less than 10 per cent) up to $\as \sim 0.8\,.$
 Due to this, as a first step, for crude quantitative estimate
 one could use one-loop Minkowskian expressions (\ref{w1}),
  (\ref{w2}), (\ref{w34}). \par
  4. Note also that a modification of the PT expansion by new
 prescription ${(\alpha_s)^k\,\to{\bf w}_k(\alpha_s)}\,$ \ for
 1-argument observable, like total cross-sections or Adler
 functions, leads -- by use of the RG algorithm --
 to a non-power APT result.

 \section{Discussion}
\begin{itemize}
\item   First, the {\it possible use of novel functions}
 $\{{\bf w}^{\rm APT}_k(\as)\}$ in pQCD has to be mentioned.
  In practice, the RG improving of PT results is limited by the
 ``1-argument" objects, like total cross-sections and D-functions.
 For the ``2-argument" ones (diffraction amplitudes, structure
 functions), one is forced to use special tricks,
 \eg, projection on 1-argument moments. \par
   New RG-inspired expansions over
 $(\as)^k\to{\bf w}^{\rm APT}_k(\as)\,$
 provide another bypass solution to this issue. In other words, we
 recommend using novel expansion functions for theoretical analysis
 of divers physical amplitudes in the  low-energy (low momentum
 transfer) regions. For a semi-quantitative quick analysis one could
 use one-loop Minkowskian functions (\ref{1min1}), (\ref{1min2-3})
 with effective values \cite{zay06} of the $\Lambda\,$ parameter. \par

 \item We believe that the observed feature of {\it interaction
 saturation} could have a rather {\it general nature}.
   Indeed, the saturation of the interaction intensity
 or, rather, its self-saturation in the ``strong coupling limit"
 could be correlated with analogous features of some soluble
 QFT models, like massless two-dimensional\cite{ThirMod58,MSh:58}
 Thirring and sine-Gordon model\cite{Aref74} equivalent
 to the massive Thirring one\cite{PR-75}. Additional evidence can
 be gathered from models with infrared fixed point, like the
 Gross-Neveu model \cite{GN:74,GN:07} and the 3-dimensional
 $\varphi^4\,$ model \cite{VKT:79}.

 \item  One of the possible ways of further analysis, to reveal
 this aspired generality, could be connected with RG study of the
 corresponding non-quantized field models by the recently
 devised\cite{KPSh:96,ShK:99-06}) {\it method of renormgroup
 symmetries}(=RGS) for boundary value problems of classical
 mathematical physics. Here, one has to find appropriate RGS
 invariants and then relate them with a quantized version with the
 help of the functional integral representation and the saddle-point
 procedure \cite{KazSh:80}.

\end{itemize}

\section*{Acknowledgements}
  The author is grateful to A.Bakulev, I. Caprini, J. Fischer,
  D.I. Kazakov, and
 S.Mikhailov for useful discussion and, especially, to I. Arefeva
 for valuable advice. The work has been supported in part by RFBR
 grant No.05-01-00992 and by Scient.School grant 5362.2006.2.
 \bigskip

\begin{appendix}
\makeatletter
 \setcounter{subsection}\z@
 \setcounter{subsubsection}\z@
 \def\thesubsection{\arabic{subsection}}%
 \def\thesubsubsection{\alph{subsubsection}}%
 \@addtoreset{equation}{section}%
 \def\theequation@prefix{\thesectionAlph}%
 \def\theequation{\theequation@prefix\arabic{equation}}%
 \def\thesectionAlph{\Alph{section}}%
 \def\thesection{\large Appendix~\Alph{section}}%
\makeatother
 \section{\sf Basic APT formulae}

 Within the Analytic Perturbation Theory, the set of common QCD
 coupling functions and its powers is changed for a nonpower set
 of ghost-free functions connected by recurrent relations. For
 instance, in the one-loop case, instead of the polynomial set
\begin{equation}\label{AI-1}
 \albars^{(1)}(Q^2)=\frac{1}{\beta_0\,\ln\left(Q^2/\Lambda^2\right)};
 \quad (\albars^{(1)}(Q^2))^2=\frac{1}{\beta_0^2\,\ln^2\,L};\quad
 (\albars^{(1)}(Q^2))^3\quad \dots \end{equation}
 one deals with
\begin{equation}\label{analE}
 \al_E^{(1)}(Q^2)\equiv\Acal_1(Q^2)=
 \frac{1}{\beta_0\,L}+\frac{\Lambda^2}{\beta_0\,(\Lambda^2-Q^2)};
 \quad \Acal_2=\frac{1}{\beta_0^2\,L^2}-\frac{\Lambda^2\,Q^2}
 {\beta_0^2\,(\Lambda^2-Q^2)^2};\quad\Acal_3;\quad\dots\eeq
  related by the differential relation
  {\small
\begin{equation}\label{recurr1}
 -\frac{1}{k}\frac{\Acalk^{(1)}(Q^2)}{\,d L}=\beta_0\,
 \Acal_{k+1}^{(1)}(Q^2);\quad L=\ln\frac{Q^2}{\Lambda^2}\,.\eeq }
 The functions $\{\Acalk(Q^2)\}$ form a basis for expansion of
 RG-invariant
 functions depending on one kinematic argument, $Q^2={\bf Q^2}-Q_0^2$,
 \ the transferred momentum squared. For example, the Adler function
 is presented there in the form of non-polynomial perturbation
 expansion
  \[ D(Q^2)=\sum_{k}^{} d_k \Acalk(Q^2).\]
  These ``Euclidean" expansion functions are related to the common
  $[\albars]^k$ ones by the prescription
  {\small
\begin{equation} \label{AAPT}
 \mathcal{A}_k(Q^2)=\int_0^\infty\frac{\rho_k(\sigma)}{\sigma
 +Q^2}\,d\sigma, \quad \rho_k(\sigma)=\frac{1}{\pi}\, \mbox{Im}
\!\left[\albars(-\sigma- i\varepsilon)\right]^k\,,\eeq}
 that provides correspondence in the weak coupling limit: \
 $\Acalk\to\as^k\,$ as \ $\as\to 0\,.$ \par
  At the same time, within the APT, one can define ghost-free
 ``Minkowskian" expansion functions for RG-invariant observables in
 another representation, for observable depending on $s\,,$ c.m.
 energy squared, like (relation of) total cross-section(s)
 \[ R(s)=\sum_{k}^{} d_k\,\Agothk(s);\quad \Agothk(s)=
 \int^{\infty}_{s}\frac{d \sigma}{\sigma}\,\rho_k(\sigma)\,. \]

  These Minkowskian functions are connected with the Euclidean ones
   by integral transformations
  \[\label{R-oper}
 \Agothk(s)=\frac{i}{2\pi}\,\int^{s+i\varepsilon}_{s-i
 \varepsilon}\frac{dz}{z}\,\mathcal{A}_k(-z)\,;\quad\Acal_k(Q^2)
  = Q^2\int^{\infty}_0\frac{\,\Agothk(s)\,d s}{(s+Q^2)^2}\,.\]

  The first of them, Minkowskian effective coupling, in the
 one-loop case has a simple form
{\small

\begin{equation}\label{1min1}
\alpha_M^{(1)}(s)=\Agoth_1^{(1)}(s)=\left.\frac{1}
 {\pi\beta_0} \arccos\frac{L_s} {\sqrt{L_s^2 +\pi^2}}
 \right|_{L_s>0}= \frac{1}{\pi\,\beta_0}\arctan\frac{\pi}{L_s}
 \,,\quad L_s=\ln \frac{s}{\Lambda^2}\,.\end{equation} }
 Accordingly,
\begin{equation}\label{1min2-3}
 \Agoth_2^{(1)}(s)=\frac{1}{\beta_0^2}\frac{1}{L^2+\pi^2};\quad
 \Agoth_3^{(1)}(s)=\frac{1}{\beta_0^3}\frac{L}{(L^2+\pi^2)^2};\dots\,.\eeq
   Quite analogously, one can devise\cite{four03,monp04} analogous
 expansion functions $\aleph_k\,$ for the ``distance picture"
 \footnote{On the base of 3-dimensional Fourier transformation
$$
\,\bar{\psi}(Q)=(2\pi)^{-2}\int d{\bf r}\psi(r)e^{i{\bf Q r}}\,.$$}
 \beq\label{aleph}
 \aleph_k(\tfrac{1}{r^2})= r\,\int^\infty_0\,d Q\,\sin(Qr)\,\Acalk(Q)=
\frac{2}{\pi}\,\int^\infty_0\frac{d\,Q}{Q}\,\sin(Qr)\,
 \int^\infty_0\frac{d \sigma\,\rho(\sigma)}{\sigma+Q^2}\,.\eeq                           

 The convenient form of the APT formalism uses a {\it spectral
 density $\rho(\sigma)\,$} taken from perturbative input (\ref{AAPT}).
 Then all the involved functions in the mentioned pictures look like
 {\small  \begin{equation}\label{kl}
 \Acalk(Q^2)=\frac{1}{\pi}\int\limits_{0}^{\infty}
\frac{\rho_k(\sigma)\,d\sigma}{\sigma+Q^2},\,\,
\mathfrak{A}_k(s)=\frac{1}{\pi}\int\limits^{\infty}_{s}
\frac{d \sigma}{\sigma}\,\rho_k(\sigma),\,\,
\aleph_k\left(\tfrac{1}{r^2}\right)=\int\limits^{\infty}_0
\frac{\rho_k(\sigma)\, d\sigma}{\sigma}\,\left(1-e^{
 -r\sqrt{\sigma}}\right).\end{equation}  }
 In the 1-loop case
 $$
 \rho^{(1)}_1=\frac{1}{\beta_0\left[L_{\sigma}^2+\pi^2
 \right]}\,; \ \ k\,\beta_0\,\rho_{k+1}^{(1)}(\sigma)=
 -\frac{d\,\rho_k^{(1)}(\sigma)}{d\,L_{\sigma}};
 \quad L_{\sigma}= \ln\frac{\sigma}{\Lambda^2}\,$$
 As it was noted above, these expressions were generalized for the
 higher-loop case with transitions across heavy quark thresholds
 and successively used (see, e.g., Ref.\cite{Sh01epjc}) for fitting
 of various data.
 \end{appendix}

  \end{document}